\documentclass[11pt,a4paper]{article}
\pdfoutput=1
\usepackage{jheppub}
\usepackage{tensor}
\usepackage{physics}
\usepackage{cancel}
\usepackage{amsmath,amsfonts}
\usepackage[english]{babel}
\usepackage{tikz}
\usepackage[normalem]{ulem}
\usetikzlibrary{3d}
\usetikzlibrary{decorations.pathmorphing}

\usepackage{amsthm}
\usepackage{slashed}
\usepackage[utf8]{inputenc}
\usepackage[T1]{fontenc}
\usepackage{mathtools}
\usepackage{float}
\usepackage{empheq}

\def\R{\mathbb{R}}

\def\cL{\mathcal{L}}

\def\p{\partial}

\def\/{\over}
\def\ov{\over}

\def\rn{\rangle}
\def\ln{\langle}

\def\s{\sigma}

\def\vphi{\varphi}
\def\a{\alpha}
\def\b{\beta}
\def\d{\delta}
\def\k{\kappa}
\def\g {\gamma}
\def\la {\lambda}
\def\w {\omega}

\def\l{\ell}
\def\mn{{\mu\nu}}
\def\rs{{\rho\sigma}}

\def\n {\nabla}
\def\L{\Lambda}
\def\D{\Delta}
\def\G {\Gamma}

\def\Tr{\mathrm{Tr}}

\def\r{\mathrm}

\def\_{\hspace{2cm}}
\def\-{\\\notag}
\def\={&=&}

\newcommand\be{\begin{equation}}
\newcommand\ee{\end{equation}}

\newcommand{\bea}{\begin{eqnarray}}
\newcommand{\eea}{\end{eqnarray}}

\newcommand{\bpm}{\begin{pmatrix}}
\newcommand{\epm}{\end{pmatrix}}

\newcommand{\bit}{\begin{itemize}}
\newcommand{\eit}{\end{itemize}}

\newcommand{\ben}{\begin{enumerate}}
\newcommand{\een}{\end{enumerate}}

\newcommand\bsp{\begin{split}}
\newcommand\esp{\end{split}}

\def\ms{\medskip}

\def\bpsi{\bar{\psi}}

\def\l{\ell}

\def\qq{\qquad}

\def\UV{\text{UV}}
\def\AdS{\text{AdS}}
\def\m{\text{m}}

\subheader{\begin{flushright}
\end{flushright}}

\title{Lessons on Eternal Traversable Wormholes in AdS}

\author[a,b]{\small Ben Freivogel,}
\author[a]{Victor Godet,}
\author[a,b]{Edward Morvan,}
\author[a]{Juan F. Pedraza}
\author[a]{and Antonio Rotundo}
\affiliation[a]{\small Institute for Theoretical Physics, University of Amsterdam, 1090 GL Amsterdam, Netherlands}
 \affiliation[b]{GRAPPA, University of Amsterdam, 1090 GL Amsterdam, Netherlands}
\emailAdd{benfreivogel@gmail.com}
\emailAdd{v.z.godet@uva.nl}
\emailAdd{edwardmorvan@gmail.com}
\emailAdd{jpedraza@uva.nl}
\emailAdd{a.f.rotundo@uva.nl}

\abstract{We attempt to construct eternal traversable wormholes connecting two asymptotically AdS regions by introducing a static coupling between their dual CFTs.
We prove that there are no semiclassical traversable wormholes with Poincar\'e invariance in the boundary directions in higher than two spacetime dimensions.
We critically examine the possibility of evading our result by coupling a large number of bulk fields.
Static, traversable wormholes with less symmetry may be possible, and could be constructed using the ingredients we develop here.}

\begin{document}

\maketitle

\section{Introduction and summary}
Recent progress  has shown that traversable wormholes are not confined to science fiction  and may even exist in the real world \cite{Gao2016TraversableWormholes,Maldacena:2018lmt,Maldacena2018TraversableWormholes,Fu:2018oaq, Bachas:2017rch}. This raises fascinating questions about the precise rules for such wormholes. Since traversable wormholes rely on violations of the Null Energy Condition (NEC), which is only possible quantum mechanically, one may expect that such wormholes are small, fragile, or short-lived, as seen in the examples that have been constructed so far.

To understand better what types of wormholes are possible, we attempt to construct static traversable wormholes in the controlled setting of asymptotically Anti-de Sitter spacetime, within a regime where the semiclassical approximation is valid. In the context of the AdS/CFT correspondence, a traversable wormhole should be dual to two conformal field theories (CFTs) which live on the two asymptotic boundaries of the spacetime.  Traversable wormholes can be constructed by introducing an appropriate coupling between these two theories, CFT$_L$ and CFT$_R$.

Bulk solutions that correspond to traversable wormholes require a violation of the NEC. This is possible in standard quantum field theory, e.g. through the Casimir effect, but a coupling between the two CFTs is unavoidable in our setting. In the decoupled system, no operator in CFT$_L$ can influence CFT$_R$, which implies that no signal can be transmitted through the bulk. The existence of a traversable wormhole solution in the decoupled system would violate this ``no-transmission principle'' \cite{Engelhardt:2015gla} which follows from basic postulates of the holographic dictionary.

Hence, we violate the NEC by introducing a double trace deformation that couples the two CFTs \cite{Aharony:2005sh} and has the form $h \hspace{1pt}  \mathcal{O}_L(x) \mathcal{O}_R(x)$. Our major assumption which makes the analysis tractable is that the solution preserves Poincar\'e invariance in the field theory directions. Assuming this symmetry, we can pick a gauge where the metric takes the form
\be\label{metricansatz}
ds^2 = a(z)^2 \left(\eta_{\mu \nu} dx^\mu dx^\nu + dz^2\right)~.
\ee
We look for solutions with two asymptotic regions, so that the `scale factor' $a(z)$ diverges at two locations while remaining nonzero in between. Furthermore, we assume that the bulk field providing the NEC violation preserves Weyl invariance. This allows us to compute explicitly the NEC violating stress-energy tensor by using a conformal map to flat spacetime.

This is a crucial simplification: in general we need to solve the Einstein equation using the expectation value of the stress-energy tensor as a source. The stress-energy tensor should be computed by doing quantum field theory in the background metric defined by $a(z)$. However, the expectation value of the stress-energy tensor $\langle T_{\mu \nu} (z) \rangle$  depends non-locally on the metric function $a(z)$, rendering the problem apparently intractable. Weyl invariance allows us to package the non-local dependence of the stress-energy tensor on the metric in terms of a single parameter encoding the `width' of the geometry.

Within these assumptions, we demonstrate a no-go result:
the effect of the double trace deformation is too small to support a semiclassical wormhole.
In order to establish this result, we consider various strategies for enhancing the NEC violation and show that they cannot work.

First, we argue that increasing the coupling does not help because the ``quantum inequalities'' \cite{Ford:1994bj} bound the amount of NEC violation. It is an interesting open problem to demonstrate a more general and more rigorous bound on NEC violation for a quantum field theory in a geometry with two asymptotically AdS regions when couplings between the boundaries are allowed.

Second, we try to add conventional matter in the bulk.  We present the detailed analysis of an additional bulk scalar field with a quartic potential, as well as establishing a general result showing that adding any additional matter satisfying the NEC does not allow for a semiclassical wormhole  with Poincar\'e invariance in the field theory directions. Our result is rigorously true when the NEC violating fields are Weyl invariant, allowing for an explicit calculation, but we suspect that adjusting the field content will not change the result.

Finally, we try to increase the number of species contributing to the NEC violation. Although this allows one to make the curvature small in Planck units, this strategy is problematic because a large number of species is believed to lower the UV cutoff as
\be
M_{\text{UV}}^{D-2} \leq {1\ov N}{M_{p}^{D-2}}\,,
\label{eq-lore}
\ee
where $N$ is the number of species and $D$ is the number of spacetime dimensions in the bulk \cite{Dvali:2007hz, Dvali:2007wp, Dvali:2008ec, Kaloper2015LargeFieldInflation}. We show that although a large number of species can reduce the curvature of the wormhole, the radius of curvature is always at or below the UV cutoff. We discuss the possibility of violating the lore \eqref{eq-lore} by choosing appropriately the field content, making use of cancellations in the one-loop renormalization of Newton's constant. However, we argue that \eqref{eq-lore} can never be softened because it would imply the existence of traversable wormholes between two asymptotically AdS boundaries without a coupling between the two CFTs, in contradiction with the ``no-transmission principle''. This also agrees with non-perturbative arguments regarding the renormalization of Newton's constant.

The failure to construct a controlled solution within our assumptions can be partially understood heuristically as follows. In the absence of the coupling between the two boundaries, the ground state is simply two unentangled CFTs in their ground state, and the corresponding geometry is simply two copies of vacuum AdS. Turning on the coupling $h \hspace{1pt}  \mathcal{O}_L(x) \mathcal{O}_R(x)$ will lead to an amount of entanglement of order the coupling $h$- in other words, the entanglement is of order one if the coupling is perturbative. On the other hand, a controlled traversable wormhole should have a smooth geometry, leading to an entanglement of order $N^2$. This heuristic argument, however, leaves open the possibility that the construction can succeed by going to strong coupling or increasing the number of fields that are coupled. Our more detailed arguments rule out these possibilities.

We have shown that there is no semiclassical solution with Poincar\'e invariance along the boundary directions and a Weyl invariant field in the non-local coupling. This suggests avenues for future constructions based on a less symmetric ansatz. One could try to import the recent construction of long-lived traversable wormholes in flat space \cite{Maldacena2018TraversableWormholes} to the AdS setting. This construction makes use of magnetic fields, which break the transverse Poincar\'e invariance. More generally, we expect that a static traversable wormhole  should look like an AdS-Schwarzschild black hole or black brane near the two asymptotic boundaries. These metrics do not preserve Poincar\'e invariance, which further motivates reducing  the amount of symmetry. We could also consider NEC violating matter that is not conformally invariant but we do not expect our results to change dramatically.

We must note that many constructions of traversable wormholes in general relativity exist in the literature but they involve either exotic matter \cite{Morris:1988cz, Morris:1988tu, Visser:1989kh, Visser:1989kg, Poisson:1995sv, Barcelo:1999hq,Visser:2003yf} or higher-derivative theories \cite{Bhawal:1992sz, Thibeault:2005ha, Arias:2010xg} which seem to lack a UV completion \cite{Camanho:2014apa}. On the other hand, introducing a coupling between two CFTs should be perfectly physical in the context of AdS/CFT.

This paper is organized as follows. In section
\ref{sec:confflatwh}, we present the computation of the stress-energy tensor generated by the non-local couping and we describe the potential wormhole solution. In section \ref{sec:challenges}, we describe challenges in obtaining a semiclassical solution. In section \ref{sec:increasing}, we argue that increasing the non-local coupling does not help. In section \ref{sec:addmatter}, we consider adding conventional matter but we prove a no-go theorem showing that this cannot lead to a semiclassical solution. In section \ref{sec:largeN}, we consider using a large number of fields in the non-local coupling but we argue that the lowering of the UV cutoff always forbids the existence of a semiclassical solution. We conclude in section \ref{sec:disc} with a discussion and some final remarks.

\section{Poincar\'e wormholes} \label{sec:confflatwh}

We look for traversable wormholes connecting two asymptotically AdS$_{d+1}$ spacetimes. In order to make the problem analytically tractable, we assume Poincar\'e invariance in the boundary directions. Using this symmetry, we can pick a gauge such that the metric takes the form
 \begin{equation}
 	\label{eq:ansatzStaticWH}
	ds^2=a^2(z)\left(-dt^2+d\vec{x}^2+ dz^2\right)\,,
\end{equation}
where $\vec{x}=(x_1,\ldots,x_{d-1})$ are boundary coordinates and $z$ is the radial coordinate in the bulk.\footnote{For later convenience we also define coordinates $x^\mu=(t,\vec{x})$ and $y^{m}=(x^\mu,z)$.} This metric is foliated by flat $\mathbb{R}^{1,d-1}$ slices and is similar to the Poincar\'e patch of AdS. The geometry is completely determined by one function, the conformal factor $a(z)$. For solutions with two asymptotically AdS boundaries, this means that $a(z)$ should have two simple poles, say at $z=\pm  \frac{L}{2}$, and be positive in the range $-\frac{L}{2} < z <  \frac{L}{2}$.

\subsection{Setup}

We consider a theory of gravity with negative cosmological constant coupled to matter,
\begin{equation}
  S=\frac{1}{16\pi G}\int d^{d+1} x\sqrt{-g}\,\left(R-2\Lambda\right)+ S_{\text{matter}}\,,
\end{equation}
where $\Lambda=-\frac{d(d-1)}{2\ell_{\text{AdS}}^2}$. In order to find traversable wormholes in such a theory, we need to violate the null energy condition (NEC) in the bulk. This is possible in the framework of semiclassical gravity, where the matter fields are treated quantum mechanically, but the geometry is kept classical. Following \cite{Gao2016TraversableWormholes}, we do this by introducing a non-local coupling between the two boundaries
\be\label{deformation}
\d S = h \int d^d x \,\phi_L(x) \phi_R(x)\,.
\ee
Here $\phi_{L,R}(x)$ corresponds to a bulk field and the subindex $L/R$ means that it is evaluated at the left/right boundary. In AdS/CFT, such a deformation is achieved by coupling together the two CFTs with a double trace operator.

In \cite{Gao2016TraversableWormholes}, the deformation (\ref{deformation}) was activated for a short time on an eternal AdS black hole, i.e. a non-traversable wormhole.
The resulting quantum stress-energy tensor made the wormhole traversable but only in a very small time window. Our work differs from \cite{Gao2016TraversableWormholes}
in two aspects. First, we start from the vacuum state, which consists of two unentangled copies of the same CFT. Second, we are interested in finding \emph{eternal} traversable wormholes so we turn on the deformation for all times.

Finally, a word on the methodology is in order. Since our ansatz (\ref{eq:ansatzStaticWH}) is conformally flat, we compute the quantum stress-energy tensor in flat spacetime and map the result to our wormhole background by means of a Weyl transformation. For this to be possible, we require $S_{\text{matter}}$ to be Weyl invariant. A simple choice is a conformally coupled scalar field. The boundary conditions for the scalar field are chosen as follows. Near the two asymptotic boundaries the behavior is
\be
\phi(z) \sim \a_\pm \left(\tfrac{L}{2}\pm z\right)^{\D_+} + \b_\pm \left(\tfrac{L}{2}\pm z\right)^{\D_-}\,,
\ee
where
\be
\D_\pm = {d \pm 1 \ov 2}\,.
\ee
which follow by performing a Weyl transformation to the flat space solution $\phi(z)\sim \a_\pm \left( {L\ov 2}\pm z\right) + \b_\pm $ near the boundaries.\footnote{This can also be derived from the formula $\D(d-\D) =- m^2\l_{AdS}^2$ since the conformal coupling $\xi R \phi^2$ with $\xi = {d-1\ov 4d}$ gives an effective mass $m^2\l_\AdS^2 = -{d^2 -1 \ov 4}$ close to the boundaries.} We choose the boundary condition
\be
\a_\pm = 0\,,
\ee
which, in the alternate quantization, implies that the dimension of the dual operator is given by $\D_-$.\footnote{In the standard quantization the dimension of the double trace term would be $2\Delta_{+}=d+1>d$ which would make the coupling irrelevant.} Upon a Weyl transformation, this condition corresponds to imposing Neumann boundary conditions on the plates in Minkowski space.

\subsection{Minkowski configuration} \label{flatspaceconfig}

We compute the stress-energy tensor by a Weyl transformation from Minkowski spacetime. In the following, we will specialize to $3+1$ dimensions. The generalization to other dimensions is described in Appendix \ref{generalD}.

The configuration in flat space consists of two infinite plates located at $z=- L/2$ and $z= L/2$, respectively, and a massless scalar field living in the region between the plates. The non-local coupling then takes the form
\be\label{nlterm}
\d S = h\int d^3 x \,\phi\left(x,-\tfrac{L}{2}\right)\phi\left(x,\tfrac{L}{2}\right)\,,
\ee
where $x = (t,x_1,x_2)$ denote the transverse coordinates. We will denote $y=(x,z)$ the coordinate of a point between the plates. The stress-energy tensor generated by the non-local coupling can be computed by point splitting
\be
\ln T_\mn(y)\rn = \lim_{y'\to y} \left( {\p \ov \p y^\mu} {\p \ov \p y'^\nu} \d G(y,y') - {1\ov 2} \eta_\mn \eta^\rs {\p \ov \p y^\rho} {\p \ov \p y'^\s}\d G(y,y') \right)\,,
\ee

where $\d G(y,y)$ is the correction to the Feynman propagator due to the non-local coupling. As explained above, we are imposing Neumann boundary conditions on the plates. Note that imposing instead Dirichlet boundary conditions would make the deformation \eqref{nlterm} vanish. The Feynman propagator with Neumann boundary conditions has a simple form in a mixed representation where we go to momentum space in the transverse directions
\be
G(x,z;x',z') = \int {d^3k\ov(2\pi)^3} e^{i k(x-x')}G_\text{mixed}(z,z'; k)\,,
\ee
where $k=(\w,k_1,k_2)$ is the momentum associated to the transverse directions $(t,x_1,x_2)$.
The propagator with Neumann boundary conditions takes the form
\be
G_\text{mixed}(z,z'; k) ={1\ov \k \,\r{sin}(\k L)} \r{cos}\!\left(\k \left(z_-+\tfrac{L}{2}\right)\right) \,\r{cos}\!\left(\k\left(z_+-\tfrac{L}{2}\right)\right), \qq
\ee
with $z_- = \r{min}(z,z')$, $z_+ = \r{max}(z,z')$ and $\k = \sqrt{\w^2- k_1^2 - k_2^2}$. We will perform the computation in  Euclidean signature where the propagator takes the form
\be
G_\text{mixed}(z,z'; k) ={1\ov |k| \,\r{sinh }(|k| L)} \r{cosh}\!\left(|k| \left(z_-+\tfrac{L}{2}\right)\right) \,\r{cosh}\!\left(|k|\left(z_+-\tfrac{L}{2}\right)\right), \qq
\ee
where $|k|=\sqrt{\w^2+k_1^2+k_2^2}$.

The correction to the two-point function due to the non-local coupling (\ref{nlterm}) is given by
\be
\d G(y,y') = h\int d^3\tilde{x}  \,G\left(\tilde{x},-\tfrac{L}{2}; y\right) G\left(y'; \tilde{x},\tfrac{L}{2}\right) + (y\leftrightarrow y')\,.
\ee
Using the mixed representation, we can rewrite this as
\be
\d G(y,y') = h \int {d^3 k \ov (2\pi)^3} {1\ov |k|^2 \,\r{sinh}^2(|k|L)} \r{cosh}\!\left(|k| \left(z + \tfrac{L}{2}\right)\right) \,\r{cosh}\!\left(|k|\left(z'-\tfrac{L}{2}\right)\right) e^{i k(x-x')} + (y\leftrightarrow y')\,.
\ee
From the above expression it can be seen that
\be
\lim_{y'\to y} \eta^\rs\p_\rho\p_\s' \d G(y,y') = 0\,.
\ee
Weyl invariance and transverse Lorentz symmetry imply that the stress-energy tensor has the form
\be\def\arraystretch{1}\arraycolsep=3pt
\langle T^{\text{flat}}_\mn \rangle =  -\rho\bpm -1 & &  &  \\  & 1 &  & \\    &  & 1 \\ & & & -3 \epm\,.
\ee
The parameter $\rho$ can in principle depend on $z$, but the conservation of the stress-energy tensor requires $\rho$ to be a constant. In order to determine this constant it suffices to compute only one of the components. For instance, we can compute
\be
\ln T^{\text{flat}}_{zz} \rn =\lim_{y'\to y} \p_z \p_z' \d G(y,y')\,.
\ee
The calculation further simplifies by going to the midpoint $z=0$, which gives
\be
\ln T^{\text{flat}}_{zz} \rn= -h \int {d^3 k \ov (2\pi)^3} {1\/2 \,\r{cosh}^2\left({|k| L\/2} \right)} = -{h\/6 L^3}\,.
\ee
Hence, the full stress-energy tensor is given by
\be\def\arraystretch{1}\arraycolsep=3pt
\langle T^{\text{flat}}_\mn \rangle = {h\/ 18 L^3}\bpm -1 & &  &  \\  & 1 &  & \\    &  & 1 \\ & & & -3 \epm\,.
\ee

\subsection{Wormhole solution}\label{sec:WHsol}

The conformal mapping allows us to compute the expectation value of the stress-energy tensor in the conformally flat geometry \eqref{eq:ansatzStaticWH}. This is given by
\be
\langle T_\mn^{\text{NL}} \rangle = {1\ov a(z)^2} \langle T^{\text{flat}}_\mn \rangle\,.
\ee
The superscript `NL' is a reminder that this component of the stress-energy tensor is generated by the non-local coupling, but generically there can be other contributions.

An important issue here is that the above Weyl transformation is anomalous in even dimensions. The anomaly generates a higher-derivative term in $\langle T_\mn^{\text{NL}} \rangle$ which prevents us from solving Einstein's equation. For the time being we will assume that the anomaly term is negligible, but we will come back to this issue in section \ref{sec:confanom}. We can also go to odd spacetime dimensions where there is no anomaly.

We define $\la$ to be the dimensionless parameter measuring the amount of negative energy generated by the non-local coupling. It is defined by
\be
8\pi \ln T_{00}^\text{flat} \rn=-{\la\ov L^4}\,,
\ee
so that $\la\sim hL$ up to a numerical factor of order one. The stress-energy tensor in the geometry \eqref{eq:ansatzStaticWH} is then given by
\be\def\arraystretch{1}\arraycolsep=3pt
8\pi \langle T_\mn^{\text{NL}} \rangle = {\la\/  a^2 L^4}\bpm -1 & &  &  \\  & 1 &  & \\    &  & 1 \\ & & & -3 \epm\,.
\ee

We will solve the semiclassical Einstein equations
\be
G_\mn - {3\/\l_\AdS^2} g_\mn = 8\pi G\ln T_\mn^\text{NL}\rn\,.
\ee
The $zz$ component can be written as
\be\label{eq:Einstein}
{1\/2}a'(z)^2 + V(a)=0\,,
\ee
which can be thought as a particle in the potential
\be\label{eq:potential}
V(a) = {G \lambda \/2 L^4}-{a^4\/2\l_{\text{AdS}}^2}\,.
\ee
The other components can be obtained from the $z$-derivative of \eqref{eq:Einstein}.

\begin{figure}[t!]
  \centering
  \includegraphics[width=0.46\textwidth]{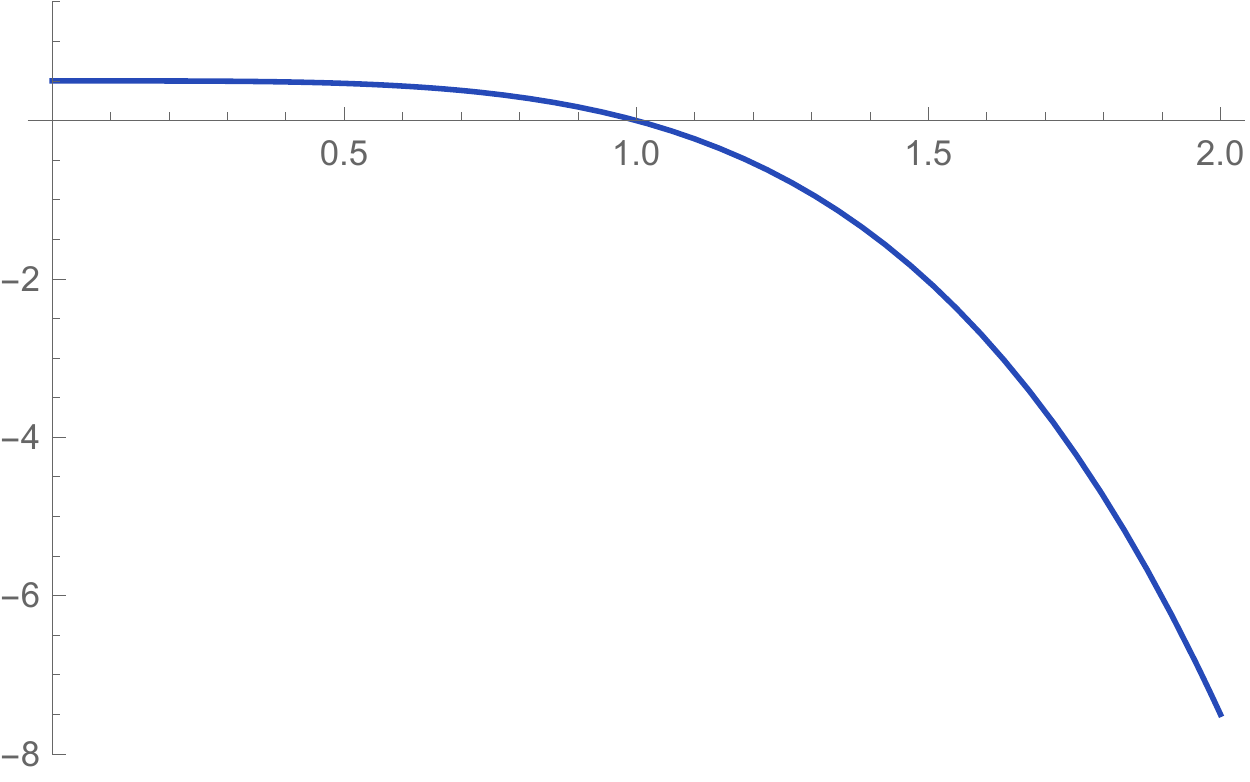}  \includegraphics[width=0.46\textwidth]{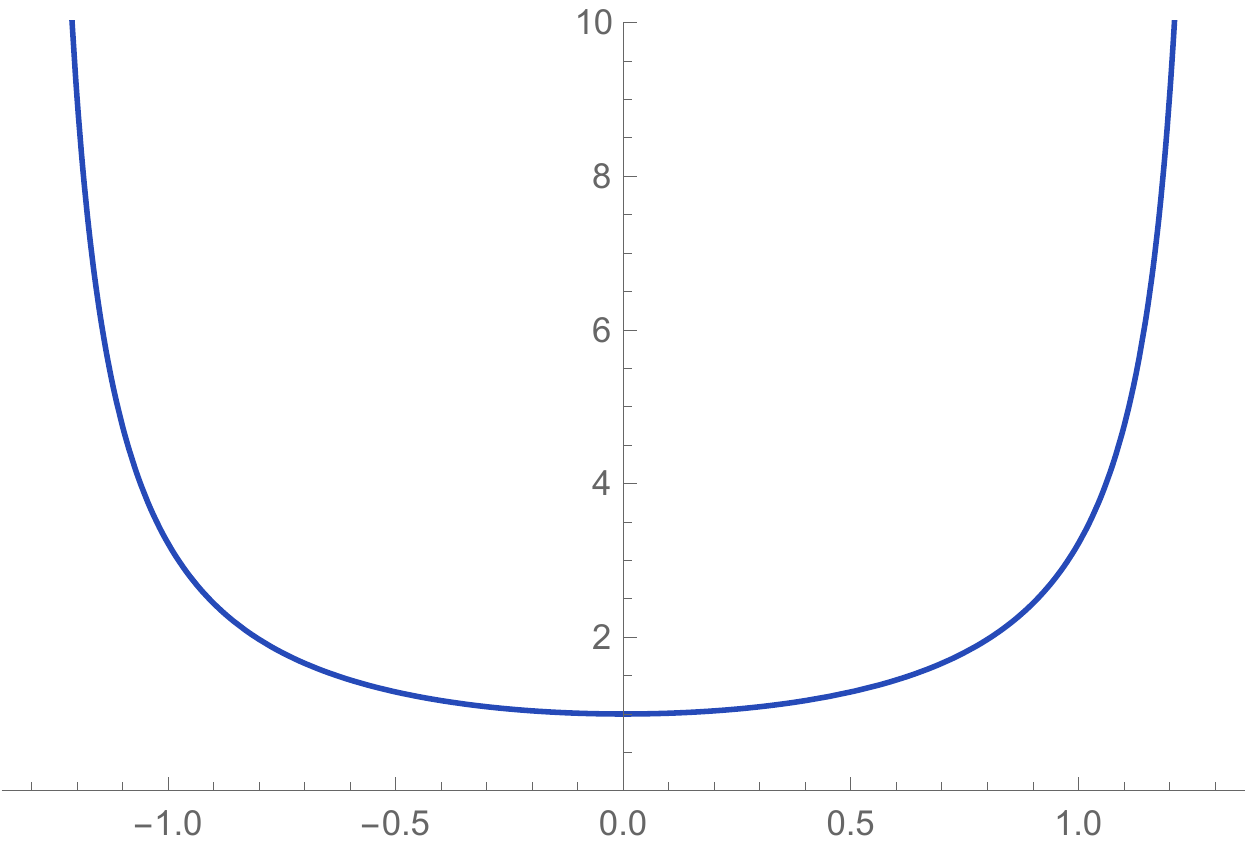}
  \begin{picture}(0,0)
\put(-392,120){{\small $V(a)$}}
\put(-99,125){{\small $a(z)$}}
\put(-215,109){{\small $a$}}
\put(-12,14){{\small $z$}}
\end{picture}
    \caption{Typical shape of the potential $V(a)$ and the conformal factor for a wormhole solution $a(z)$. For the plots we have set $\ell_{\text{AdS}}=1$ and we have set $a(0)=1$.}
 \label{fig:solution}
\end{figure}
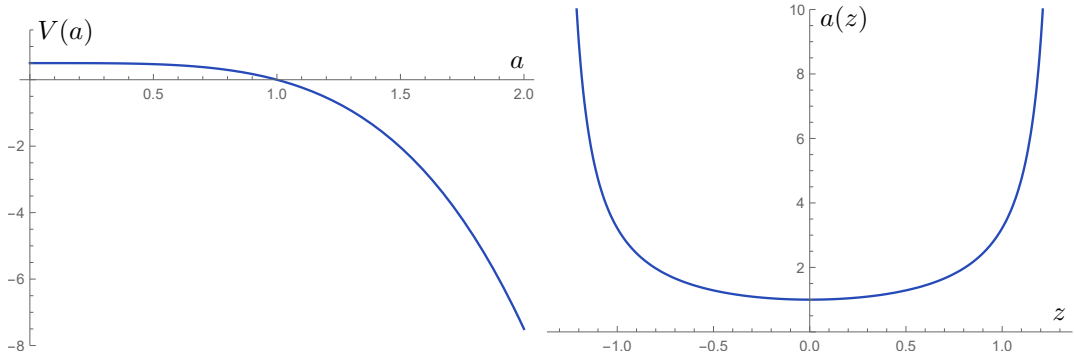

There is a diffeomorphism
\be
y^\mu\to\zeta\, y^\mu\,,\qquad L\to \zeta\,L\,,\qquad h\to \zeta^{-1} h\,,\qquad a(z)\to \zeta^{-1} a(z)\,,
\ee
which allows us to set $a(0)=1$. Furthermore, we focus on solutions with reflection symmetry so we have $a'(0)=0$ and can restrict the domain of integration to $z\geq 0$. Evaluating \eqref{eq:Einstein} at $z=0$ gives us the relation
\be\label{eq:const}
\frac{1}{\ell^2_{\text{AdS}}} = \frac{G \la }{L^4}\,,
\ee
This allows us to eliminate $\la$ in \eqref{eq:Einstein} and to determine the value for the range $L$
\be
{L\/2} = \int_0^{L/2} dz = \int_1^{\infty} \frac{da}{a'} = \l_\AdS \int_1^\infty {da\/\sqrt{a^4-1}} \,.
\ee
Performing the integral gives
\be\label{relLtolAdS}
L = c\,\l_\AdS, \qq c={2\sqrt{\pi}\G(\tfrac{5}{4})\/\G(\tfrac{3}{4})}  \sim 2.62\,.
\ee
This defines a one-parameter family of wormhole solutions. The potential (\ref{eq:potential}) and a typical wormhole solution are shown in Figure \ref{fig:solution}.

\section{Challenges} \label{sec:challenges}

\subsection{Planckian curvature}

The most important issue of our solutions is that they generically have large curvatures.
Combining the equations \eqref{eq:const} and \eqref{relLtolAdS} implies that
\begin{equation}\label{planckiancurv}
\left( {\l_{\text{AdS}}\ov\l_p} \right)^2 \sim  \la \,,
\end{equation}
where $\l_p$ is the Planck length and $\sim$ means proportionality up to an order one numerical factor. Since the computation of the stress-energy tensor is valid only in the perturbative regime, $\la \ll 1$, this leads to a wormhole with super-Planckian curvature. Hence, the wormhole solutions described in the previous section are outside the regime where semiclassical gravity can be trusted. In the following sections we will explore various potential ideas to try to resolve this issue, but we first discuss two more challenges.

\subsection{Weyl anomaly\label{sec:confanom}}

We have used a Weyl transformation from flat space to compute the stress-energy tensor in the geometry \eqref{eq:ansatzStaticWH}. For this to be possible, we used a Weyl invariant field in the non-local coupling. However, Weyl invariance can be anomalous at the quantum level. The anomaly is problematic in our setup because, as we will see below, it can be of the same order as the effect that we use to support the wormholes.

In four dimensions, there is always an anomaly. The stress-energy tensor on the conformally flat metric \eqref{eq:ansatzStaticWH} is related to the stress-energy tensor in flat space by \cite{Birrell:1982ix}
\be\label{anomT}
\ln T_\mn\rn = {1\/a^2}\ln T_\mn^\text{flat}\rn + {1\/16\pi^2} \left(2 \a F_\mn -{\b\/9} H_\mn \right)\,.
\ee
The anomalous piece is the second term and it is expressed in terms of four-derivative terms
\bea
F_\mn \= R_{\mu}^{~\rho} R_{\rho\nu} -{2\/3} R R_\mn -{1\/2} R_\rs R^\rs g_\mn +{1\/4}R^2 g_\mn\,, \\
H_\mn \= 2 R_{;\mn} - 2 g_\mn \Box R -{1\/2} g_\mn R^2 + 2 R R_\mn\,.
\eea
The rational numbers $\a$ and $\b$ can be extracted from the tables of \cite{Birrell:1982ix}. For massless free fields, they are given by
\bea\label{anomalyalpha}
\a \= {1\/360} (n_S + 11 n_F + 62 n_V)\,,\\
\b \={1\/20} (n_S(1-5\xi) + n_F + 2 n_V)\,,
\eea
with $n_S$ conformally coupled scalars, $n_F$ Dirac fermions and $n_V$ vectors. The parameter $\xi$ is the coupling to $R$ which should be taken to be $\xi = {1\/6}$ because we want a Weyl invariant theory. From the above expression, we can see that $\a$ is strictly positive. In fact, the trace of \eqref{anomT} shows that $\a$ is the $a$-anomaly of the 4d theory which does not vanish for a unitary theory \cite{Cardy:1988cwa, Komargodski:2011vj, Casini:2017vbe}. This implies that there is no way to make the anomaly vanish in four dimensions.

The main problem with the anomalous piece is that it contains four-derivative terms which prevent us from solving Einstein's equation. For a given field, the size of the anomaly is of the order of
\be
|T_\mn^{\text{anomaly}}| \sim {1\/\l_{\text{AdS}}^4}\,.
\ee
In the configuration with a single non-locally coupled field, the anomaly can be ignored in the regime $L < \l_\AdS$ because we have $\la \ll 1$. This condition is already necessary as will be derived later in \eqref{hierarchies}. In the configuration where we have a large number of fields, discussed in section \ref{sec:largeN}, the anomaly dominates over the cosmological constant term which prevents us from obtaining a semiclassical solution.

In odd dimensions, there is no Weyl anomaly: the Weyl transformation of the stress-energy tensor does not contain an anomalous piece. Thus, the stress-energy tensor in the wormhole spacetime \eqref{eq:ansatzStaticWH} can be obtained from the stress-energy tensor in flat space by the classical formula
\be
\ln T_\mn\rn = {1\/a^2}\ln T_\mn^\text{flat}\rn\,.
\ee
Hence, when using a large number of fields, we will focus only on odd dimensions.

\subsection{Casimir energy}\label{sec:Casimir}

There is another problem for building eternal traversable wormholes between asymptotically AdS regions. Negative energy can already be present in the wormhole geometry without the need to turn on a non-local coupling. Indeed, in the flat space configuration described in section \ref{flatspaceconfig}, the Casimir effect \cite{Milton:2001yy} generates a stress-energy tensor of the form \cite{Birrell:1982ix}
\be\def\arraystretch{1}\arraycolsep=3pt
T^{\text{Casimir}}_\mn \sim {1\/ L^4}\bpm -1 & &  &  \\  & 1 &  & \\    &  & 1 \\ & & & -3 \epm\,.
\ee
 This negative energy is more important than the one generated by the non-local coupling, which is multiplied by $\la \ll 1$. Hence, it seems that the non-local coupling is unnecessary! However, it should not be possible to build a semiclassical wormhole between two asymptotically AdS geometries without coupling the two dual CFTs. This would violate the ``no-transmission principle''  \cite{Engelhardt:2015gla} because signals could be sent from one asymptotic boundary to the other without any coupling between the two CFTs. We note that the sign of the Casimir energy can be modified by changing the boundary conditions of the fields. However, having a positive Casimir energy is also a problem because it would overwhelm the non-local coupling.

The issue of Casimir energy should be present in any attempt to build eternal traversable wormholes. In the asymptotically AdS$_2$ version \cite{Maldacena:2018lmt} (see also \cite{Bak:2018txn, Bak:2019mjd}), this is avoided because the Weyl anomaly precisely cancels the Casimir energy. As explained there, this is enforced by $SL(2,\R)$ invariance.  In higher dimensions, the wormhole is less symmetric which makes such a cancellation unlikely. From the analysis of section \ref{sec:confanom}, we see that this cancellation happens neither in four dimensions nor in odd dimensions where the Weyl anomaly vanishes. Thus, if our conformally flat wormholes are to be consistent, some mechanism has to ensure that the Casimir energy is negligible. For example, a supersymmetric spectrum with supersymmetric boundary conditions leads to a vanishing Casimir energy. This holds despite the fact that the wormhole geometry breaks supersymmetry\footnote{This can be checked explicitly by showing that the geometry \eqref{eq:ansatzStaticWH} does not have a covariantly constant spinor, except when it is flat space or Poincar\'e-AdS.} as can be seen by making a Weyl transformation from the flat space configuration.

\section{Increasing the non-local coupling} \label{sec:increasing}

We might hope that a solution with $\l_{\text{AdS}}$ large in Planck units can be obtained in the strong coupling regime $\la\gg 1$. In fact, we will show that increasing the coupling cannot lead to a very large negative energy. This can be done by adapting the ``quantum inequalities'' \cite{Ford:1994bj}. The authors proved that for any state $|\psi\rn$ of a free massless scalar in Minkowski spacetime, there is a bound
\be
\hat{\rho} \geq - {c\/ t_0^4}\,,
\ee
where $\hat\rho$ is the energy density averaged over a time interval of characteristic length $t_0$,
\be
\hat{\rho} \equiv  \int_{-\infty}^{+\infty} dt \,f(t) \ln \psi | T_{00} |\psi\rn\,,
\ee
and $f$ is a smearing function which determines the number $c$. This shows that the smeared energy cannot get ``too negative''. In their proof, the smearing function is a Lorentzian but the same argument can be repeated for a more general smearing function as long as its Fourier transform decays sufficiently fast. This is because the bound is proportional to the integral
\be\label{boundintegral}
\int_0^\infty d\w\, \w^3 \hat{f}(\w)\,,
\ee
where $\hat{f}(\w)$ is the Fourier transform of $f$. In particular, it is possible to obtain a bound when $f$ being compactly supported. This follows from a Theorem by Ingham \cite{doi:10.1112/jlms/s1-9.1.29} which determines how fast the Fourier transform of a compactly supported function can decay. This theorem guarantees that there are compactly supported functions whose Fourier transform decays exponentially, e.g. as $e^{-|\w|^{1/2}}$, which is fast enough to make \eqref{boundintegral} converge.

In our Minkowski configuration, we can consider a causal diamond centered at $z=0$ which is as large as possible without touching the plates. Because the diamond is not in contact with the plates, the quantum state inside this causal diamond is that of a free massless scalar field. Hence, we expect that the quantum inequalities should be applicable if the smearing function is supported in this diamond. This is possible by taking a compactly supported function on a time interval of length $L$. The resulting bound is
\be
\ln T_{00} \rn \gtrsim -{1\/ L^4}\,,
\ee
up to an order one numerical factor. Thus, the best we can achieve by increasing the coupling would lead to a wormhole with Planckian curvature
\be
\left({\l_{\text{AdS}} \ov \l_p}\right)^2 \sim 1\,.
\ee
This shows that increasing the non-local coupling does not help in making the wormhole semiclassical.

\section{Adding conventional matter}\label{sec:addmatter}

In the previous sections, we have shown that the negative energy generated by the non-local coupling is too small to support the wormhole. From \eqref{eq:const}, we can see that to have $\lambda\ll 1$ we need a hierarchy of scales $\l_p\ll L\ll \ell_{\text{AdS}}$.
However, with only the non-local coupling and the cosmological constant we have $L\sim \ell_{\text{AdS}}$ as shown in \eqref{relLtolAdS}. We can attempt to solve this problem by adding a new \textit{classical} source in the Einstein equation. This will introduce a new scale which in principle could be used to separate $L$ from $\ell_{\text{AdS}}$ or remove the necessity of this hierarchy altogether. In this section, we prove a no-go theorem showing that this is not possible: adding matter that satisfies the null energy condition cannot make the wormhole semiclassical.

\subsection{Scalar field}

Before going to the general situation, we consider a bulk scalar field minimally coupled to gravity, described by the action
\begin{equation}
	S_\m=-\int{\sqrt{-g}\,d^{4}x\left(\frac{1}{2} g^\mn \partial_{\mu}\phi\partial_{\nu}\phi +V(\phi)\right)}\,.
\end{equation}
We assume that the matter preserves Poincar\'e invariance in the transverse directions so that $\phi$ depends only on $z$. In our geometry \eqref{eq:ansatzStaticWH}, the matter stress-energy tensor is
\begin{equation}
\label{eq:scalarStressEnergy}
\begin{split}
	T_{zz}^\m&=a^{2}\left(\frac{\phi'^{2}}{2a^{2}}-V\right)\,,\\
	T_{xx}^\m&=T_{yy}^\m=-T_{tt}^\m=-a^{2}\left(\frac{\phi'^{2}}{2a^{2}}+V\right)\,.
\end{split}
\end{equation}
The scalar field does not violate the NEC at classical level because $T_{zz}^\m+T_{tt}^\m=\phi'^{2} \geq 0$. Its equation of motion is
\begin{equation}
	2a'\phi'+a\phi''=a^{3}\,\partial_{\phi}V\,.
\end{equation}
Einstein equation gives
\begin{equation}
\label{eq:eomGen}
\begin{split}
	&a'^{2}=-\frac{G \la}{ L^{4}}+\frac{8\pi G}{3}a^{4}\left(\frac{\phi'^{2}}{2a^{2}}-V\right)\,,\\
	&\frac{a''}{a^{3}}=-\frac{8\pi G}{3}\left(\frac{\phi'^{2}}{2a^{2}}+2V\right)\,.
\end{split}
\end{equation}
The term generated by the non-local coupling is the negative term proportional to $\la$ in the first equation. We can explicitly see that this term is necessary by considering the expression
\be\label{eq:necCond}
\left({a'\/a^2} \right)' = { 2G\la \/ L^4 a^3 } - 4\pi G{\phi'^2\/a}\,.
\ee
In a wormhole solution, $a'/a^{2}$ goes from $0$ at the throat, to $1/\l_\AdS$ at the boundary. Therefore, its derivative needs to be positive somewhere, implying that $\la$ cannot be zero. More explicitly, integrating the equation between $z=0$ and $z= L/2$ gives
\begin{equation}
	\frac{1}{\l_\AdS}= \int_{0}^{ L/2} dz \left({a'\/a^2} \right)' =\int_{0}^{ L/2} dz{\left(\frac{2G\lambda}{a^{3} L^{4}}-4\pi G\frac{\phi'^{2}}{a}\right)}\leq \frac{G\lambda}{ L^{3}}\,.
\end{equation}
This leads to the following lower bound
\begin{equation}
	\lambda \ge  \left({\l_\AdS\/\l_p}\right)^2\left(\frac{ L}{\l_\AdS}\right)^{3}\,.
\end{equation}
We are in a regime where $\la \ll 1$ which implies that we must have $ L\ll \l_\AdS$. We see that the scalar field does not modify the required hierarchy we pointed out at the beginning of the section. This is a special case of a more general statement \eqref{hierarchies} which will be described later. Note that this requirement guarantees that the Weyl anomaly is negligible, as discussed in section \ref{sec:confanom}.

\begin{figure}[t!]
\centering
	\includegraphics[width=0.48\textwidth]{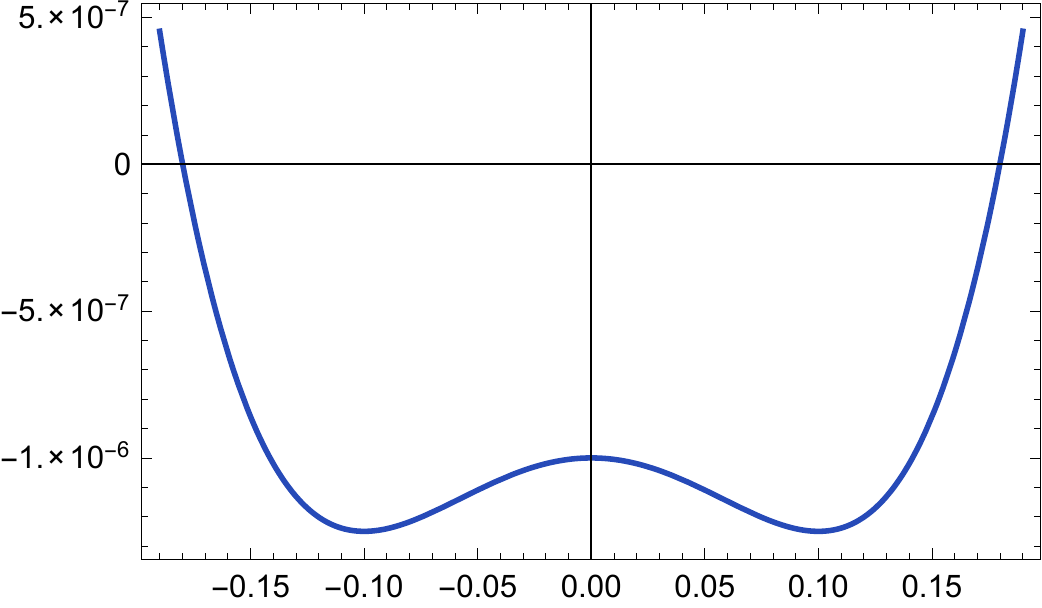}
	\includegraphics[width=0.47\textwidth]{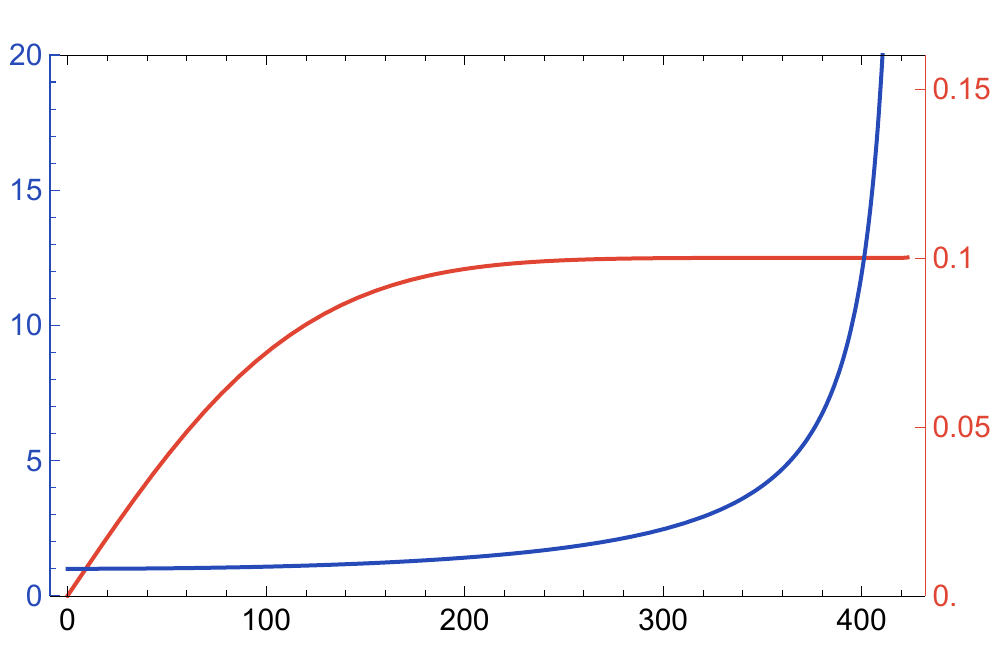}
	\begin{picture}(0,0)
  		\put(-76,38){{$a(z)$}}
		\put(-170,70){{$\phi(z)$}}
		\put(-9,8){{$z$}}
		\put(-310,128){{$V(\phi)$}}
		\put(-407,85){{$\phi$}}
	\end{picture}
\caption{\textbf{Left:} Shape of the potential  $V(\phi)$. \textbf{Right:}  The corresponding solution. We are using Planck units for the two plots.}
\label{fig:phi4PotSol}
\end{figure}
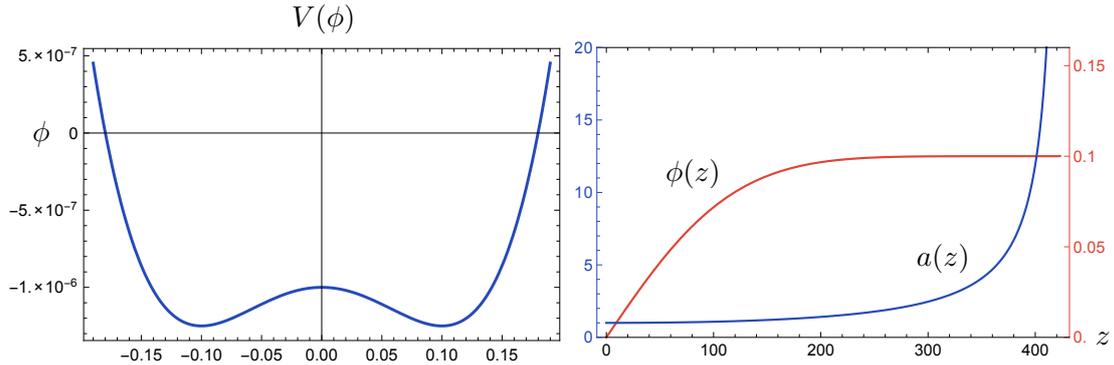

We solve numerically the coupled ODEs for $a(z)$ and $\phi(z)$ using Mathematica. The solution is obtained by integrating the second order Einstein equation which leads to more stable numerics.\footnote{The two Einstein equations in \eqref{eq:eomGen} are not independent. We can obtain the second order differential equation by taking the derivative of first one and using the $\phi$ equation of motion to remove the $\phi''$ terms.} We consider a Higgs-like potential
\begin{equation}
	V=-{m_{0}^2 \/2} \phi^2+{c^2 \/4}\phi^4+V_{0}\,.
\end{equation}
which is illustrated in Figure \ref{fig:phi4PotSol}.

We look for solutions of $\phi$ that interpolate between the two minima of $V$ at the two asymptotically AdS regions. These solutions are odd so we can restrict the range of integration to $ 0\leq z\leq {L\/2}$. We consider the following boundary conditions
\begin{equation}
\begin{split}
	&a(0)=1\,,\quad a'(0)=0\,;\\
	&\phi(0)=0\,,\quad \phi(\tfrac{L}{2})=\phi_{L}\,,
\end{split}
\end{equation}
where $\phi_L$ is the value corresponding to the minimum of the potential.\footnote{The value of $L$ is determined dynamically because it corresponds to location at which $a(z)$ diverges. For this reason, imposing the condition at the boundary is a bit tricky. In practice we impose a second boundary condition at the throat $\phi'(0)=\phi'_0$. The correct value of $\phi'_0$ is determined through algorithmically to ensure that $\phi$ approaches the right value at $z=L/2$.}

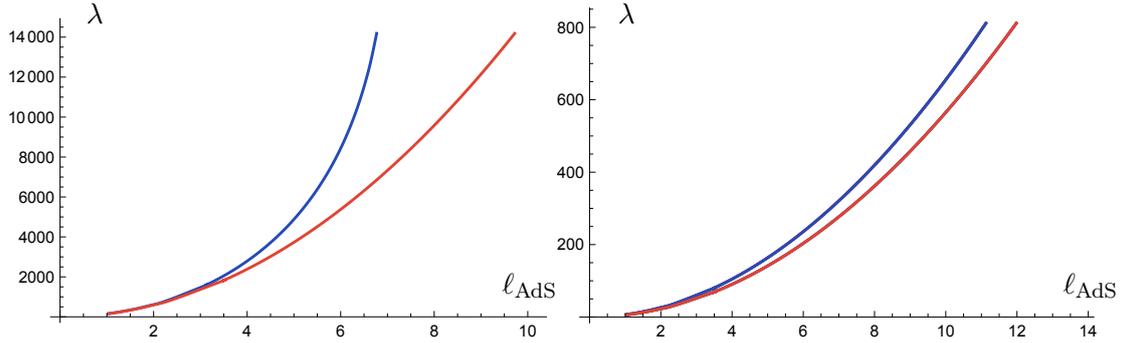
\begin{figure}[t!]
\centering
\includegraphics[width=0.47\textwidth]{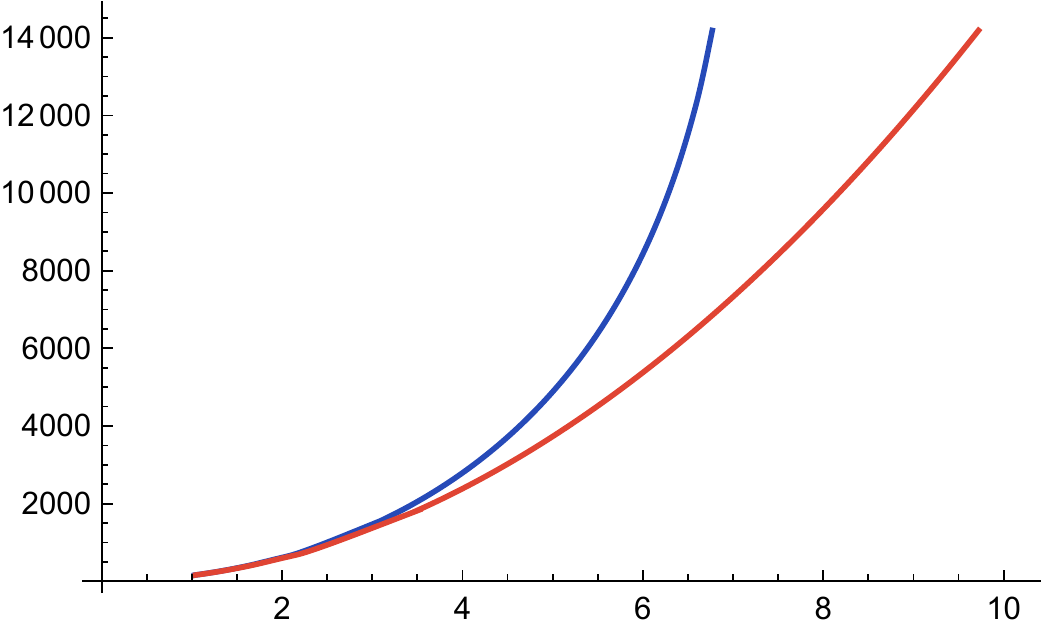}
\includegraphics[width=0.47\textwidth]{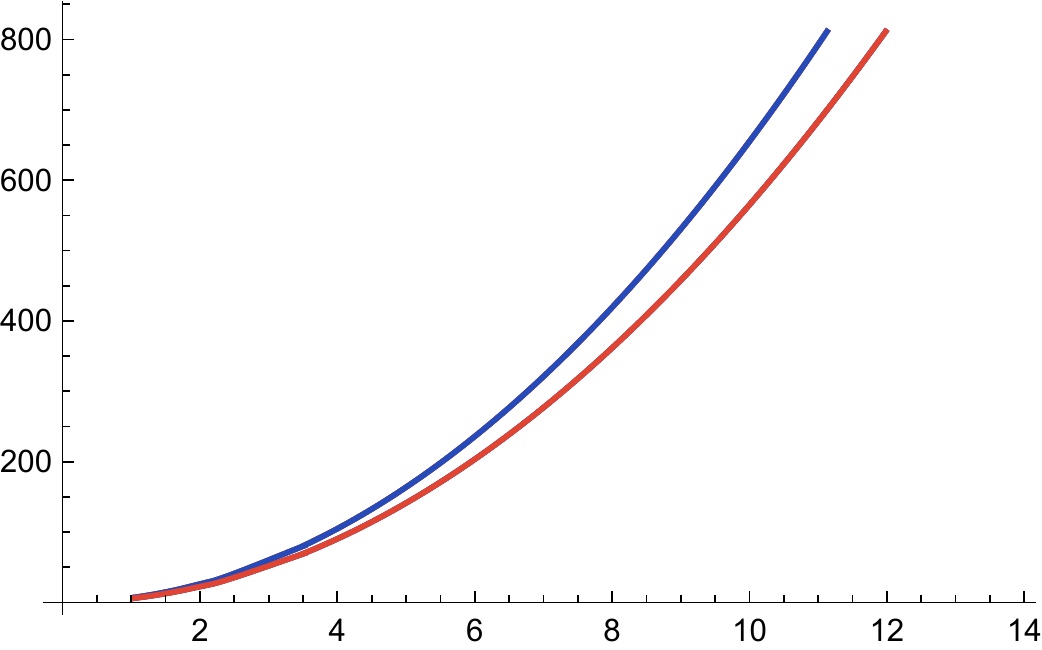}
\begin{picture}(0,0)
  \put(-20,18){{ $\ell_{\text{AdS}}$}}
  \put(-230,18){{ $\ell_{\text{AdS}}$}}
  \put(-382,120){{$\lambda$}}
  \put(-183,120){{$\lambda$}}
\end{picture}
\caption{We plot $\lambda$ as a function of $\ell_{\text{AdS}}$ while varying $V_0$ in the interval $ -1\leq V_0\leq -10^{-4}$ and we use Planck units. The different curves correspond to different choices for the other parameters in the potential. \textbf{Left:} $m_0=c=10^{-1}$ (blue), $m_0=c=10^{-2}$ (red). \textbf{Right:} $c=10^{-i}$ with $i=2,3$ and $m_0=10^{-1}c$ (blue) and $m_0=10^{-2}c$ (red). The series corresponding to different values of $i$ are indistinguishable. It's impossible to have $\la \ll 1$ if we want $\l_\AdS$ to be large in Planck units.}
\label{fig:lambdaOverL}
\end{figure}

From the numerical solutions, we can find the value of $\lambda$ and $\ell_{\text{AdS}}$ using
\begin{equation}
	\lambda = {8\pi L^4 \/3} \left(\frac{\phi'(0)^2}{2}-V(\phi (0))\right)\,,\quad
	\ell_{\text{AdS}}=\sqrt{-\frac{3 }{8 \pi  G V(\phi_L)}}\,.
\end{equation}
An example of solution is given in Figure \ref{fig:phi4PotSol}.
In general we notice that $\lambda$ can be made small only at the cost of making $\l_\AdS$ small in Planck units. In Figure \ref{fig:lambdaOverL}, we plot $\lambda$ as a function of $\l_\AdS$ for a large sample of parameters. We only keep solutions which leads to $\l_\AdS>1$. In all cases, even for $\l_\AdS$ close to $\l_p$ we do not find solutions consistent with $\lambda\ll1$. This means that the addition of the bulk scalar field does not help in making the wormhole semiclassical.

\subsection{No-go theorem}

We will now show that any kind of conventional matter in the bulk does not help in making the wormhole semiclassical, assuming that the matter respects Poincar\'e invariance in the transverse directions. We can model the addition of bulk matter by the addition of a term $f(a)$ in Einstein equation
\begin{equation}
  \label{eq:modEinsEq}
  a'^2={a^{4}\/\l_{\AdS}^2}-\frac{G \la }{ L^{4}}+{f(a)\/\l_\m^2}\,,
\end{equation}
where $\l_m$ is a characteristic length scale of the additional matter. We show below that $f(1)$ needs to be positive so $\l_\m$ can be chosen so that $f(1)=1$. We are also using conventions in which $a(0)=1$. Evaluating the equation at $z=0$ gives
\be
\la = {L^4\/ G}\left( {1\ov \l_{\AdS}^2}+{1\/\l_\m^2}\right)\,.
\ee
From the above formula, we see that $\la \ll 1$ implies that
\be\label{hierarchies}
{L\/\l_{\AdS}} \ll {\l_p\/L} \qq \text{and} \qq {L\/\l_m} \ll {\l_p\/L}\,.
\ee
The Einstein equation can be rewritten
\be\label{einsteinrewritten}
a'^2 = {f(a)- f(1)\/\l_m^2} + {a^4-1\/\l_{\AdS}^2}\,.
\ee
We require the asymptotically AdS boundary condition
\be
a(z)\big|_{z\to L/2}\sim\frac{\l_{\AdS}}{ {L\/2}-z}\,.
\ee
In other words, the cosmological constant should dominate close to the boundary $z={L\/2}$. For the additional matter to be helpful, we would like to dominate close to the wormhole throat $z=0$. We can define the transition point $z_\ast$ and the corresponding conformal factor $a_*\equiv a(z_*)$ by
\be
{f(a_*)- f(1)\/\l_m^2} = {a_*^4-1\/\l_{\AdS}^2}\,.
\ee
We assume that the additional matter dominates below $z_\ast$ while the cosmological constant dominates above $z_\ast$.
Hence, we have
\begin{align}\label{throateq}
{a^4-1\/\l_{\AdS}^2} &\leq {f(a)- f(1)\/\l_\m^2}\,,  & \kern-6em 0 \leq z \leq z_\ast\,, \\
{a^4-1\/\l_{\AdS}^2} &\geq   {f(a)- f(1)\/\l_\m^2}\,,  & \kern-6em z_\ast \leq z \leq {L\/2}\,.
\end{align}
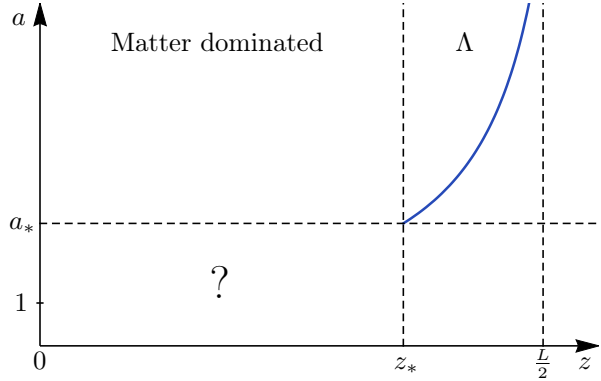
\begin{figure}[t!]
  \centering
  \includegraphics[width=0.5\textwidth]{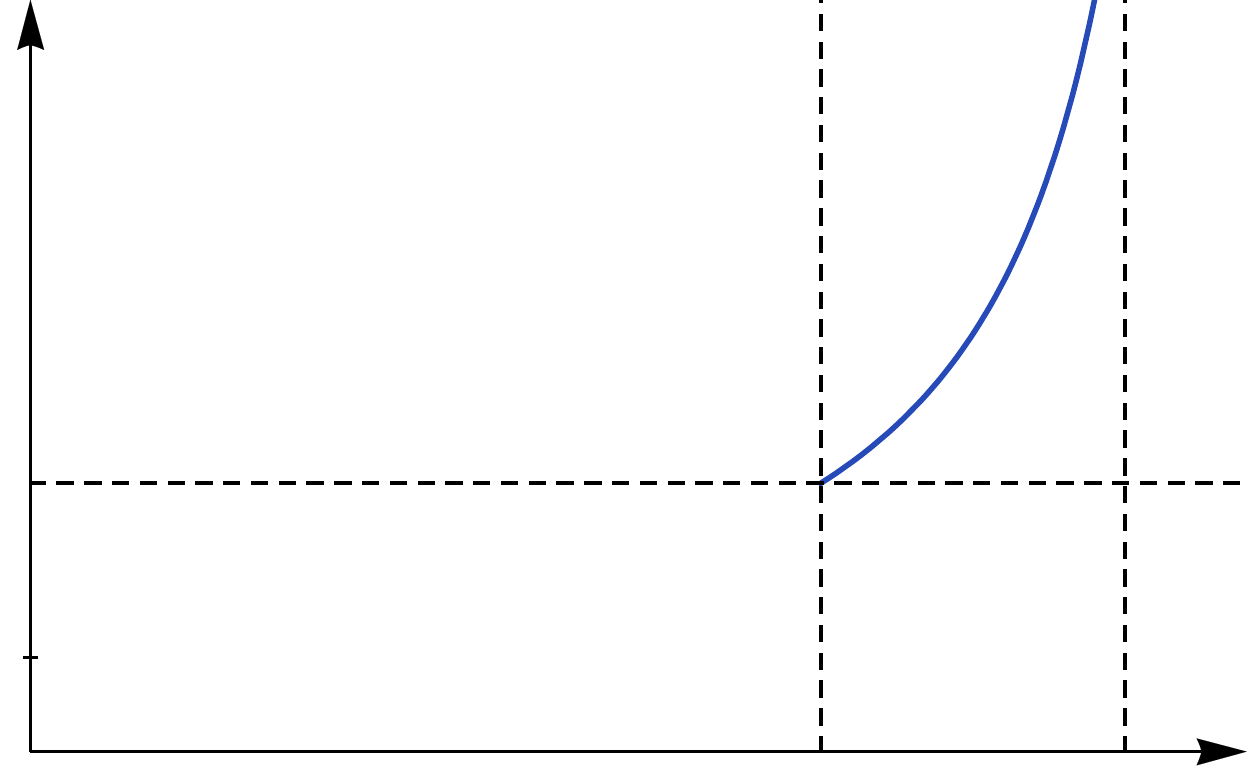}
    \begin{picture}(0,0)
  \put(-224,125){{\small $a$}}
\put(-225,48){{\small $a_*$}}
\put(-223,17){{\small $1$}}
\put(-216,-5){{\small $0$}}
\put(-81,-5){{\small $z_*$}}
\put(-29,-5){{\footnotesize $\frac{L}{2}$}}
\put(-12,-5){{\small $z$}}
\put(-150,22){{\LARGE ?}}
\put(-187,115){{\small Matter dominated}}
\put(-58,115){{\small $\Lambda$}}
\end{picture}
  \caption{We assume that the new term in the Einstein equation dominates up to some $z=z_{*}$, for which we have $a(z_{*})=a_*$. After this value the cosmological constant dominates.}
    \label{fig:WHbehaviour}
\end{figure}
We also assume that $a(z)$ is monotonically increasing close to $z=0$.\footnote{Relaxing these two assumptions cannot help in making the wormhole more semiclassical. Indeed, the above discussion shows that in order to be useful, the conventional matter needs to make $L$ as small as possible. It can be seen that relaxing these assumptions will only make things worse.} In Figure \ref{fig:WHbehaviour} we show a schematic representation of the two regimes described above.

\paragraph{Null energy condition.}

We impose the null energy condition on the additional matter.  The $zz$-component can be directly read off from \eqref{eq:modEinsEq}
\begin{equation}
  8\pi G \,T_{zz}^\m={3 \/a^2} {f(a)\/\l_\m^2}\,.
\end{equation}
We consider matter that respects the transverse Lorentz symmetry. This implies that the stress-energy tensor is diagonal and satisfies $T_{xx}^\m=T_{yy}^\m =-T_{tt}^\m$. Its conservation then implies that
\be
{d\/d z}T_{zz}^\m  + {a'\/a} T_{zz}^\m + {3 a'\/a} T_{tt}^\m = 0\,.
\ee
The resulting stress-energy tensor is
\begin{equation*}
  8\pi G \,T_{\mu\nu}^\m={1\/a^2\l_\m^2}\left(
  \begin{matrix}
      f(a)-af'(a) & 0 & 0 & 0\\
    0 & af'(a)-f(a) & 0 & 0\\
    0 & 0 & af'(a)-f(a) & 0\\
    0 & 0 & 0 & 3f(a)
  \end{matrix}\right)\,.
\end{equation*}
The null energy condition applied to the vector $\p_t+\p_z$ implies
\begin{equation}\label{NECeq}
  f'(a)\leq \frac{4 f(a)}{a}\,.
\end{equation}
In particular, this implies that $f(1) > 0$. Indeed, dividing \eqref{throateq} by $a-1$ and taking the limit $a\to 1$  implies that $f'(1) > {4 \l_\m^2 / \l_\AdS^2}$.

\paragraph{Contradiction.}

First, integrating the NEC gives
\be
f(a) \leq  a^4\,.
\ee
Intuitively, this means that the fastest function of $a$ which satisfies the NEC is a cosmological constant. Since our original problem was that the cosmological constant does not grow fast enough, no other conventional matter should help us in making the wormhole more semiclassical.

More precisely, from \eqref{einsteinrewritten} and \eqref{throateq}, we see that
\begin{align}
a'^2 & \leq {2(f(a)-1)\/\l_\m^2}\,,  &  \kern-6em  0 \leq z \leq z_\ast\,, \\ \label{conteq2}
a'^2 & \leq {2(a^4-1)\/\l_\AdS^2}\,, &  \kern-6em   z_\ast \leq z \leq {L\/2}\,.
\end{align}
Integrating the first equation implies that
\be\label{zasteq}
z_\ast = \int_{1}^{a_*} {da\/a'} \geq \frac{\l_\m }{\sqrt{2}}\int_1^{a_*} {da\/\sqrt{f(a)-1}} \geq\frac{\l_\m }{\sqrt{2}}\int_1^{a_*} {da\/\sqrt{a^4-1}}\,.
\ee
Next, we can obtain a bound on $a_*$ by integrating the second equation,
\be
{L\ov 2} - z_\ast= \int_{a_*}^{\infty} {da \/a'} \geq {\l_\AdS \ov \sqrt{2}}\int_{a_*}^{\infty} {da \/\sqrt{a^4-1}} \geq {\l_\AdS \ov \sqrt{2}}\int_{a_*}^{\infty} {da \/a^2} = {\l_\AdS \ov \sqrt{2}} {1\ov a_*}\,.
\ee
This gives
\be
a_* \geq {\sqrt{2} \l_{\AdS} \/L} \gg 1\,,
\ee
where the second inequality follows from \eqref{hierarchies}. Then, equation \eqref{zasteq} gives $z_\ast > O(1)\l_\m$ where $O(1)$ is an order one number (which is bigger than $\sim 0.5$ already for $a_* = 2$). Using again \eqref{hierarchies}, this implies
\be
z_\ast \gg L/2\,,
\ee
which is a contradiction. This shows that adding conventional matter does not help in making the wormhole semiclassical.

\section{Coupling a large number of fields}\label{sec:largeN}

In the previous section we have shown quite generally that we cannot build a traversable wormhole with just a perturbative amount of negative energy, assuming Poincar\'e invariance in the boundary directions. Increasing the coupling or adding conventional matter do not help. Another strategy is to use a large number $N$ of fields in the non-local coupling to enhance its effect. This approach was exploited in previous constructions of traversable wormholes \cite{Maldacena:2017axo, Maldacena2018TraversableWormholes, Maldacena:2018lmt}. For our wormholes, the estimate \eqref{planckiancurv} gets replaced by
\be
\left({\l_{\text{AdS}}\ov \l_p} \right)^{D-2} \sim N \la\,,
\ee
where $N$ is the number of fields and $D=d+1$ is the number of spacetime dimensions in the bulk. Thus, it seems that by taking $N$ large enough, we can obtain a large AdS radius while keeping $\la \ll 1$. However, a large number of fields implies a lowering of the UV cutoff of the theory, which can be interpreted as the renormalization of Newton's constant. On general grounds \cite{Dvali:2007hz, Dvali:2007wp, Dvali:2008ec, Kaloper2015LargeFieldInflation}, we expect that
\be\label{uvcutoff}
M_\UV^{D-2} \leq {1\ov N} M_p^{D-2}\,.
\ee
This implies that the solution cannot be made semiclassical
\be
\left({\l_{\text{AdS}}\ov \l_\UV} \right)^{D-2} \sim \la \ll 1\,.
\ee

\subsection{Renormalization of Newton's constant}

The perturbative renormalization of Newton's constant can be computed from the one-loop effective action. We will use the heat kernel expansion which provides a canonical way to regulate the divergences \cite{Vassilevich:2003xt}. In $D$ dimensions, the effective Lagrangian at one-loop is
\bea
\cL_{\text{eff}}
\=  - {1\/D} {a_0(x)\/\l_\UV^D} - {1\/D-2} {a_2(x) \/ \l_\UV^{D-2}} + \dots\,,
\eea
where $\l_\UV$ is a UV cutoff and $a_{n}(x)$ are the so-called Seeley-DeWitt coefficients which are obtained from the heat kernel expansion. The term $a_0(x)$ gives a renormalization of the cosmological constant and the term $a_2(x)$ gives the renormalization of Newton's constant.

\ms

Because of the Weyl anomaly discussed in section \ref{sec:confanom}, we will focus on odd-dimensional theories. We want a Weyl invariant theory so we consider $n_S$ massless scalars with the conformal coupling $\xi = {D-2\/ 4(D-1)}$ and $n_F$ Dirac fermions. The computation is detailed in the Appendix \ref{App:HK} and gives
\be
(4\pi)^{D/2} a_2(x)  = \left({4-D\/12 (D-1)} n_S + {1\/12} 2^{\lfloor {D/2}\rfloor} n_F\right)R\,.
\ee
In special cases of interest, this is
\be \label{a2alld}
(4\pi)^{D/2} a_2(x)=
\begin{cases}
{1\/24}(n_S+ 4  n_F) R\,, & D=3\,, \\
{1\/48}(-n_S+ 16  n_F)R\,, \qq & D=5\,, \\
{1\/24}(-n_S+ 16 n_F)R\,, \qq & D=7\,.
\end{cases}
\ee
We see that in 5d and 7d, we can make $a_2(x)$ vanish by choosing appropriately the field content. Note that this is possible because of the negative contribution from the conformally coupled scalars. In such cases, the perturbative lowering of the UV cutoff will be given by two-loop diagrams. This leads to a softer lowering of the UV cutoff, \eqref{uvcutoff} becomes
\be
M_\UV^{D-2} \leq {1\/\sqrt{N}} M_p^{D-2}\,.
\ee
Assuming that no other effects lower the UV cutoff, we obtain
\be\label{uvcutoffbetter}
\left({\l_{\text{AdS}}\ov \l_\UV} \right)^{D-2} \sim \sqrt{N} \la\,.
\ee
From this analysis, it seems that taking large enough $N$ allows for semiclassical wormholes. However, we argue in the next section that non-perturbatively, the UV cutoff is always lowered according to \eqref{uvcutoff}, preventing the possibility of having a semiclassical wormhole in this way. For related discussions on the renormalization of Newton's constant and the meaning of the negative contribution, we refer to \cite{Fursaev:1994ea, Larsen:1995ax, Donnelly:2012st, Donnelly:2014fua,Donnelly:2015hxa,Solodukhin:2015hma}.

\subsection{Non-perturbative considerations}

Non-perturbative arguments based on black hole physics suggest that the UV cutoff of a gravity theory with $N$ species is always lowered according to \eqref{uvcutoff}. These arguments are based on the rate of black hole evaporation, or entropy bounds  \cite{Dvali:2007hz, Dvali:2007wp, Kaloper2015LargeFieldInflation}. These arguments suggest that the one-loop cancellations in \eqref{a2alld} are not sufficient to lower the cutoff at the non-perturbative level.

Another consideration is the ``no-transmission principle'' which implies that a traversable wormhole between asymptotically AdS regions supported only by Casimir energy is inconsistent. Indeed, we should not be able to send signals from one asymptotic boundary to another if the two dual CFTs are decoupled \cite{Engelhardt:2015gla}.

For example, let us consider a theory in 5d with a large number of scalar and spinor fields such that $n_S=16 n_F$. This is chosen so that $a_2(x) = 0$ so that according to the above computation, Newton's constant is not renormalized at one-loop. Assuming that the UV cutoff is not lowered by other effects, \eqref{uvcutoffbetter} would imply that we can have a semiclassical wormhole by taking the number of fields large enough.

If the above scenario is really possible, we could also construct a traversable wormhole without the non-local coupling, using only Casimir energy. For this to be true, we need to make sure that the Casimir energy is non-zero and negative. Let us consider the above setup with $n_S=16 n_F$ but without a non-local coupling. We are free to change the boundary conditions of the fields because the computation of $a_2(x)$ is insensitive to them (up to boundary terms which are irrelevant here). In particular, we can choose the boundary conditions so that the $n_F$ spinors and the $4n_F$ scalar fields are in a supersymmetric configuration in the flat space setup. As discussed in section \ref{sec:Casimir}, this implies that the Casimir energy of these fields will compensate. The remaining $12 n_F$ scalar fields can be chosen to have Dirichlet boundary conditions in flat space. The Casimir energy of 5d massless scalars between two plates with Dirichlet boundary conditions is computed in \cite{Milton:2001yy} and is indeed negative. Following the discussion in \ref{sec:Casimir}, this would give a traversable wormhole supported only by Casimir energy. More generally, we expect that it should always be possible to make the Casimir energy negative by choosing adequate boundary conditions.

Thus, we obtain a configuration where a traversable wormhole connects two decoupled CFTs which is inconsistent because of the ``no-transmission principle'' of AdS/CFT. This strongly suggests, in agreement with the black hole arguments, that the non-perturbative UV cutoff is still lowered according to \eqref{uvcutoff} despite the perturbative cancellations at one loop. The wormhole cannot be made semiclassical by using a large number of fields.

\section{Discussion\label{sec:disc}}

In this paper we have investigated the possibility of  constructing eternal traversable wormholes connecting two asymptotically AdS regions by coupling the two dual CFTs. We focused on gravity solutions preserving Poincar\'e invariance along the field theory directions and used a Weyl invariant field in the non-local coupling.

Under these assumptions, the stress-energy tensor can be computed analytically. Although it violates the null energy condition, it does not provide enough negative energy to support a semiclassical wormhole. As argued from the ``quantum inequalities'' \cite{Ford:1994bj}, increasing the coupling does not help. We also proved a no-go theorem saying that adding conventional matter in the bulk cannot make the wormhole semiclassical. Another strategy is to use a large number of fields in the non-local coupling. This increases the negative energy but lowers the UV cutoff by a compensating amount, disallowing any semiclassical traversable wormholes. A one-loop computation suggests that this lowering of the UV cutoff, interpreted as a renormalization of Newton's constant, can be soften by adequately choosing the field content. However, non-perturbative arguments suggest that this cannot work \cite{Dvali:2007hz, Dvali:2007wp, Kaloper2015LargeFieldInflation}. In particular, this would lead to a traversable wormhole solely supported by Casimir energy, without a non-local coupling. This contradicts the ``no-transmission principle'' which follows from basic postulates of the AdS/CFT duality \cite{Engelhardt:2015gla}.

This argument suggests that any mechanism that enhances the effect of the non-local coupling should always make the Casimir energy negligible, as to prevent the possibility of a semiclassical wormhole without a non-local coupling. We expect this requirement to provide some guidance in the construction of eternal traversable wormholes in AdS/CFT.

There are many avenues for future research. We can consider changing the conformal dimensions of the field, going away from the conformally coupled case. The effect becomes more difficult to compute but the numerics in \cite{Gao2016TraversableWormholes} suggests that the negative energy can be increased in this way. We could also investigate situations with less symmetry. This would provide more room to produce the large hierarchy between the small quantum effect induced by the non-local coupling and the large semiclassical geometry. Adding rotation has been shown to enhance the effect of the non-local coupling \cite{Caceres:2018ehr}. Also, it should be possible to import in AdS/CFT the recent construction of long-lived traversable wormhole in Minkowski spacetime \cite{Maldacena2018TraversableWormholes}.

\section*{Acknowledgements}

It is a pleasure to thank Yang An, Elena C\'aceres, Alejandra Castro, Willy Fischler, Dami\'an Galante, Diego Hofman, Juan Maldacena and Dora Nikolakopoulou, for useful discussions. BF and AR are supported  by the ERC Consolidator Grant QUANTIVIOL.
JFP is supported by the Netherlands Organization for Scientific Research (NWO) under the VENI grant No. 680-47-456/1486. This work is part of the $\Delta$-ITP consortium, a program of the NWO that is funded by the Dutch Ministry of Education, Culture and Science (OCW).

\appendix

\addtocontents{toc}{\protect\setcounter{tocdepth}{1}}

\section{General dimensions}\label{generalD}

\subsection{Setup}

The computation done in section \ref{flatspaceconfig} can be generalized to any dimension. In $D$ spacetime dimensions, the $zz$ component of the stress-energy tensor is
\bea
\ln T^{\text{flat}}_{zz} \rn\= -h \int {d^{D-1} k \ov (2\pi)^{D-1}} {1\/2 \,\r{cosh}^2\left({|k| L\/2} \right)}\,, \-
\= - {h\/L^{D-1}} {\r{vol}(S^{D-2})\/(2\pi)^{D-1}} \int_0^{+\infty} {x^{D-2} dx \/2 \, \r{cosh}^2\left(\tfrac{x}{2} \right)}\,.
\eea
The full stress-energy tensor has the general form
\be\arraycolsep=3pt\def\arraystretch{1}
\ln T_\mn^\text{flat}\rn \sim {\la \/L^D} \bpm -1 & & &  &  \\  & 1 & &  \\ & & \ddots & & \\  & & & 1 &  \\ &  &  & & 1-D\epm\,,
\ee
where $\la \sim h L$ is the perturbative parameter and $\sim$ means up to an order one numerical factor.

After the conformal transformation to the metric \eqref{eq:ansatzStaticWH}, the Einstein equation has the same form  \eqref{eq:Einstein} as in 4d but with the potential
\be
V(a) = { G \la\/2 L^D} - {a^4\/2\l_{\text{AdS}}^2}\,.
\ee
We can redo the computation done in section \ref{sec:WHsol} and we obtain
\begin{equation}
\left( {\l_{\text{AdS}}\ov\l_p} \right)^{D-2} \sim  \la \,.
\end{equation}
We are in the perturbative regime $\la \ll 1$ so the wormhole cannot be semiclassical.

\subsection{No-go theorem}

The no-go theorem presented in the main text can be generalized to any dimension.  In $D$ dimensions, the Einstein equation takes the form
\be
a'^2 = -{ G\la\/L^D} + {a^4\/\l_\AdS^2}\,.
\ee
We consider a modified Einstein equation
\bea
a'^2 \=-{ G\la\/L^D} + {f(a)\/\l_\m^2} + {a^4\/\l_\AdS^2}\,, \-
\= {a^4 - 1 \/\l_\AdS^2} + {f(a)-1\/\l_\m^2}\,,
\eea
where $f(a)$ is subject to the same assumptions as in the main text.
We will show that $f(1)$ is positive which allows us to fix $\l_\m$ by requiring that $f(1)=1$. We also use conventions where $a(0)=1$. Evaluating Einstein equation at $z=0$ leads to
\be
\la \sim {L^d\/ G }\left( {1\/\l_\AdS^2} + {1\/\l_\m^2} \right)\,.
\ee
The null energy condition is obtained as in the four dimensional case. We first compute the stress-energy tensor corresponding to the new term we added in the Einstein equation. The $zz$ component can be read off the Einstein equation
\be
T_{zz}^\m = {(D-1)(D-2)\/16\pi G}{ f(a)\/ a^2 \l_\m^2}\,.
\ee
We can determine $T_{tt}^\m$ by solving at the conservation equation $\nabla_{\mu}T^{\mn}=0$. This gives
\begin{equation}
	\partial _{z}T_{zz}^\m+\left( D-3\right) \frac {a'} {a}T_{zz}^\m+(D-1)\frac {a'} {a}T_{tt}^\m=0\,.
\end{equation}
From this equation, we can determine
\be
T_{tt}^\m = {(D-2)\/16\pi G}{ (5-D)f(a) - a f'(a) \/ a^2 \l_\m^2}\,.
\ee
Evaluating the NEC, we obtain as in four dimensions
\begin{equation}
	T_{zz}^\m+T_{tt}^\m \geq 0 \implies  4f\left( a\right) -f'\left( a\right) a \geq 0\,.
\end{equation}
This bound implies that $f(1)$ is positive and leads to
\be
f(a)\le a^{4}\,.
\ee
The remainder  of the proof is unchanged.

\section{Heat kernel expansion} \label{App:HK}

We consider fluctuations of quantum fields around a given classical background. The effective action can be written
\be
S_{\text{eff}} = S_0 + S_{\text{1-loop}}\,,
\ee
where $S_0$ is the action of the classical background. The effective action is computed by a Euclidean path integral
\be
e^{-S_{\text{eff}}} = e^{-S_0} \int D\phi \, e^{- \phi \L\phi}\,,
\ee
where $\L$ is the operator of quadratic fluctuations. The heat kernel expansion \cite{Vassilevich:2003xt} provides a way to regularize and compute the effective action. The term that renormalizes $G_N$ is
\be
(4\pi)^2 a_2(x) = {1\/6}  \Tr (6 E +R)\,,
\ee
where the trace is over the components of the fields and $E$ is defined by
\be
-\L = g^\mn D_\mu D_\nu +E\,,
\ee
and $D_\mu = \n_\mu + \w_\mu$ is a suitable covariant derivative. We compute below $a_2(x)$ for fields of interest in $d$ spacetime dimensions. These results can also be found in \cite{Kabat:1995eq} except that the conformal coupling is not considered there.

\paragraph{Scalar.}

We consider a massless scalar field. The Lagrangian is $\cL = (\p\vphi)^2+\xi R\phi^2$. This gives $E =  -\xi R$. Hence,
\bea\label{scalar}
(4\pi)^{d/2} a_2^{\text{scalar}}(x) \=  \left({1\/6}-\xi\right)R\,.
\eea

\paragraph{Fermion.}

We consider a Dirac spinor. The Lagrangian is $\cL = \bpsi \g^\mu D_\mu \psi$. The fermionic fluctuation operator is thus $\g^\mu D_\mu$. This is a first order operator so we apply the heat kernel to its square. The identity $(\g^\mu D_\mu)^2 = g^\mn \n_\mu \n_\nu -{1\/4}R$ implies that $E = -{1\/4}R$. This gives
\bea
(4\pi)^{d/2} a_2^{\text{spinor}}(x) \={1\/12} 2^{\lfloor {d/2}\rfloor}\,.
\eea

\paragraph{Maxwell vector.}

The Lagrangian is $\cL = -{1\/2}F_\mn F^\mn = - D^\mu a^\nu D_\mu a_\nu + D^\nu a^\mu D_\mu a_\nu$. We integrate the two terms by parts and swap the two derivatives in the second term to obtain
 \be
\cL = a^\nu \Box a_\nu - a^\nu R_\mn a^\mu - (D^\mu a_\mu)^2\,.
 \ee
As usual, the last term is removed by adding a gauge-fixing term $\cL_{\text{g.f.}} = (D^\mu a_\mu)^2$. This introduces two scalar ghosts which are minimally coupled. The contribution of these ghosts is two times the one written in (\ref{scalar}) with  $\xi=0$ and an overall minus sign due to the opposite statistics. For the gauge field, we obtain $E = -R_\mn$. This gives
\bea
(4\pi)^{d/2} a_2^{\text{vector}}(x) \= {d-8\/6} R\,.
\eea

\bibliography{bibliography}

\end{document}